\documentclass{article}

\usepackage[T1]{fontenc}
\usepackage[utf8]{inputenc}
\usepackage{amsmath,cite,url}
\usepackage{graphicx}
\usepackage{color}
\usepackage{amsmath,amssymb,amsfonts}
\usepackage{algorithmic}
\usepackage{siunitx}
\usepackage{booktabs}
\usepackage{subcaption}
\usepackage{multirow}
\usepackage{tabularx}
\usepackage{enumitem}
\usepackage[colorinlistoftodos,prependcaption,textsize=tiny]{todonotes}
\usepackage{tikz}
\usepackage{floatrow}
\usepackage[dblblindworkshop,preprint]{neurips_2025}
\workshoptitle{The First Workshop on Generative and Protective AI for Content Creation}

\usepackage{glossaries}
\newacronym{cnn}{\textsc{cnn}}{Convolutional Neural Network}
\newacronym{ssl}{\textsc{ssl}}{Self-Supervised Learning}
\newacronym{sasv}{\textsc{sasv}}{Spoofing-Aware Speaker Verification}
\newacronym{svdd}{\textsc{svdd}}{Singing Voice Deepfake Detection}
\newacronym{svs}{\textsc{svs}}{Singing Voice Synthesis}
\newacronym{svc}{\textsc{svc}}{Singing Voice Conversion}
\newacronym{poi}{\textsc{poi}}{Person-of-Interest}
\newacronym{Private}{\textsc{private}}{Private dataset}
\newacronym{vad}{\textsc{vad}}{Voice Activity Detector}
\newacronym{tdnn}{\textsc{tdnn}}{Time Delay Neural Network}
\newacronym{roc}{\textsc{roc}}{Receiver Operating Characteristic}
\newacronym{auc}{\textsc{auc}}{Area Under the Curve}
\newacronym{eer}{\textsc{eer}}{Equal Error Rate}
\newacronym{fpr}{\textsc{fpr}}{False Positive Rate}
\newacronym{fnr}{\textsc{fnr}}{False Negative Rate}

\newcommand{\artisttwenty}{\textsc{artist\oldstylenums{20}}}
\newcommand{\ctrsvdd}{\textsc{ctrsvdd}}
\newcommand{\wildsvdd}{\textsc{wildsvdd}}

\def\checkmark{\tikz\fill[scale=0.4](0,.35) -- (.25,0) -- (1,.7) -- (.25,.15) -- cycle;} 

 \usepackage[dblblindworkshop]{neurips_2025}

\usepackage[utf8]{inputenc} 
\usepackage[T1]{fontenc}    
\usepackage{hyperref}       
\usepackage{url}            
\usepackage{booktabs}       
\usepackage{amsfonts}       
\usepackage{nicefrac}       
\usepackage{microtype}      
\usepackage{xcolor}         

\title{Not All Deepfakes Are Created Equal: Triaging Audio Forgeries for Robust Deepfake Singer Identification}

\author{%
  Davide Salvi \\
  Universal Music Group\\
  London, United Kingdom \\
  \And
  Hendrik Vincent Koops\thanks{Corresponding author. Accepted for presentation at the NeurIPS 2025 Workshop on Generative and Protective AI for Content Creation (non-archival)} \\
  Universal Music Group\\
  London, United Kingdom \\
  \And
  Elio Quinton \\
  Universal Music Group\\
  London, United Kingdom \\
}

\begin{document}

\maketitle

\begin{abstract}
The proliferation of highly realistic singing voice deepfakes presents a significant challenge to protecting artist likeness and content authenticity. Automatic singer identification in vocal deepfakes is a promising avenue for artists and rights holders to defend against unauthorized use of their voice, but remains an open research problem. 
Based on the premise that the most harmful deepfakes are those of the highest quality, we introduce a two-stage pipeline to identify a singer's vocal likeness. It first employs a discriminator model to filter out low-quality forgeries that fail to accurately reproduce vocal likeness. A subsequent model, trained exclusively on authentic recordings, identifies the singer in the remaining high-quality deepfakes and authentic audio. Experiments show that this system consistently outperforms existing baselines on both authentic and synthetic content.
\end{abstract}

\section{Introduction}
\label{sec:intro}
Recent advances in singing voice cloning technology enable the generation of "deepfakes" that are virtually indistinguishable from authentic recordings, posing a significant threat to content authenticity and artist likeness protection. 
In the same way that audio fingerprinting protects recordings from unauthorized use \cite{sonnleitner2015robust, wang2003industrial, joren2014panako}, we propose a system to protect a singer's vocal likeness. Our goal is to be able to identify a singer's voice in both authentic and deepfake recordings, providing a tool for artists and rights holders to defend against unauthorized uses.

The scientific community has largely pursued two separate countermeasures: deepfake detection, which aims to distinguish synthetic from authentic audio \cite{xie2024fsd, guragain2024speech, zang2024singfake, gohari2025audio}, and singer identification, which verifies a vocalist's identity \cite{mesaros2007singer, sharma2019importance}. However, identifying a singer within a deepfake remains a challenge \cite{desblancs2024real}. This paper investigates singer identification across authentic and synthetic signals.

We argue that the potential for harm from a deepfake correlates with its quality: a highly realistic fake is more dangerous than a poor one where the singer is unrecognizable. Based on this observation, we introduce a two-stage pipeline designed for maximum effectiveness against the most threatening deepfakes. First, a discriminator model filters out low-quality forgeries, which do not faithfully reproduce the vocal likeness. Second, a singer identification model trained only on authentic recordings matches the test recording to known vocal likenesses. Our experiments show that our system consistently outperforms existing baselines across both authentic and deepfake content.

\begin{figure*}
  \centering
  \includegraphics[trim={14cm 3.5cm 12cm 3.5cm},clip, width=0.7\textwidth]{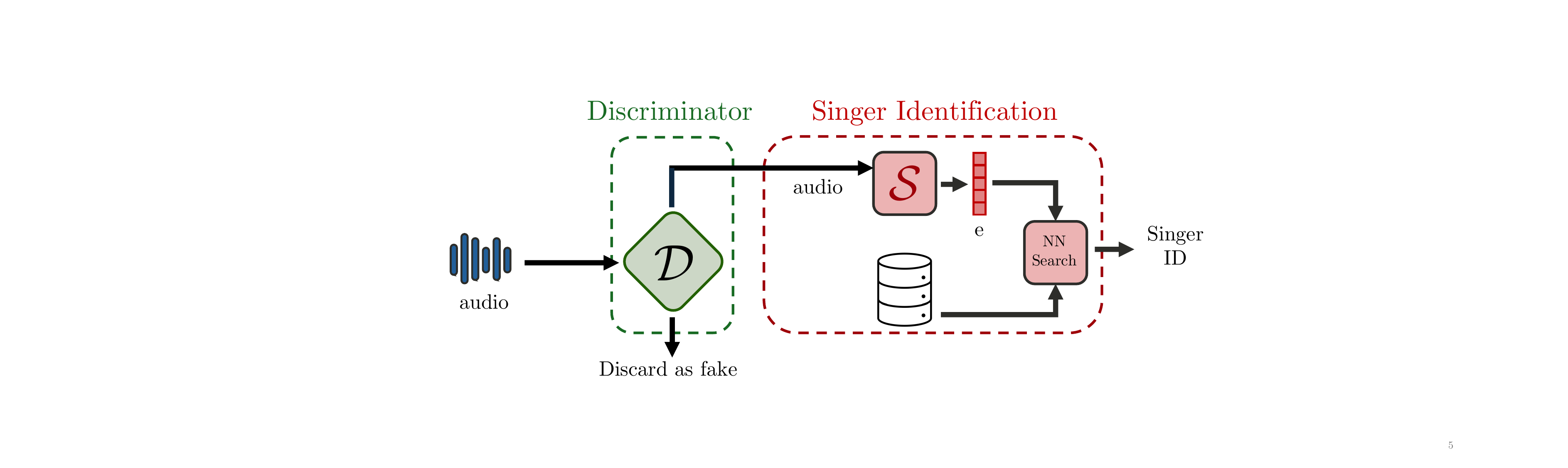}
  \caption{Proposed two-stage pipeline for singer identification. Stage 1 ($\mathcal{D}$) filters low-quality deepfakes. Tracks classified as authentic proceed to Stage 2, where singer identity is determined by nearest neighbor search using cosine distance of extracted embeddings ($e$).}
  \label{fig:pipeline}
\end{figure*}

\section{Method}
\label{sec:method}
Motivated by the notion that the potential harm that can be caused to an artist by an unauthorized deepfake is proportional to its perceptual quality, we propose a two-steps pipeline, depicted in Figure~\ref{fig:pipeline}. 
First, the recording under test is processed by a discriminator $\mathcal{D}$, which  objective is to filter out poor quality deepfakes. Because they typically exhibits significant artifacts and do not faithfully replicate the singer's vocal likeness, their potential for harm is comparatively much lower than that of higher quality deepfakes, and they are likely to be easier to detect. 
Recordings that are deemed either high quality deepfakes or authentic by $\mathcal{D}$ are passed to the second stage, where a singer identification model $\mathcal{S}$ extracts a vocalist likeness embedding to be compared with a database of known vocalist likeness embeddings. 

For the discriminator ($\mathcal{D}$) we use a Light Convolutional Neural Network (\textsc{lcnn})~\cite{lavrentyeva2019stc, wu2018light}, a compact and efficient architecture introduced for spoofing detection. It is trained to predict whether a track is a deepfake or authentic on the \textsc{ctrsvdd} dataset. Because we expect that poor quality deepfakes should be easy to detect because they features salient and unnatural artifacts, we purposefully keep $\mathcal{D}$ light and its training regime simple. Our expectation is that $\mathcal{D}$ would be effective at detecting poor deepfakes but would be fooled by high quality ones.

For the singer identification model ($\mathcal{S}$) we apply the \textsc{ecapa-tdnn} architecture~\cite{desplanques2020ecapa} to singing voice data and train it as a multi-class classifier. We selected this model architecture, shown to perform well in speech applications, on the basis that our task also focuses on human voice. 
In addition, we consider two baselines. 1- \textsc{ssl}: A Self-Supervised Learning baseline using a Siamese EfficientNet-B0 encoder, specifically designed to learn singer representations from mel-spectrograms~\cite{torres2023singer}. 2- \textsc{resnet-tdnn}: A hybrid model~\cite{VILLALBA2020101026} pre-trained on the VoxCeleb speech dataset~\cite{speechbrain} to evaluate speech-to-singing speaker identification transferability. $\mathcal{S}$ and \textsc{ssl} are trained only on authentic content. Again, this is an important property since it makes the method easier to operate and scale, and paired authentic and deepfake content is rarely available in practice.

\begin{table*}[b]
    \centering
    \resizebox{\textwidth}{!}{ 
    \begin{tabular}{llccccccr}
        ~ & ~ & ~ & ~ & \multicolumn{2}{c}{$\mathcal{D}$} & \multicolumn{2}{c}{$\mathcal{S}$} & ~ \\
        ~ & ~ & \multicolumn{2}{c}{Authenticity} & \multicolumn{2}{c}{Discriminator} & \multicolumn{2}{c}{Singer Identification} & ~ \\
        \cmidrule(lr){3-4} \cmidrule(lr){5-6} \cmidrule(lr){7-8} 
        Dataset           & Audio Contents & Authentic & Deepfake     & Training  & Testing & Training & Testing & Reference\\
        \hline
        \textsc{private}  & Full mix & \checkmark & $\times$         &  $\times$ & \checkmark & \checkmark & \checkmark & --- \\
        \artisttwenty{}   & Full mix & \checkmark & $\times$         &  $\times$ & \checkmark & $\times$ & \checkmark & Ellis in \cite{ellis2007classifying} \\
        \ctrsvdd{}        & Vocals only  & \checkmark & \checkmark   &  \checkmark & \checkmark &  $\times$ & \checkmark &  Zang et al. in \cite{zang24_interspeech} \\
        \wildsvdd{}       & Full mix & \checkmark & \checkmark  &  $\times$ & \checkmark &  $\times$ & \checkmark & Zhang et al. in \cite{zhang2024svdd} \\
    \end{tabular}
    }
    \caption{Overview of used datasets. \textit{``Full mix''} refers to a mix of vocals with (background) music. \textit{``Authenticity''} refers to bona fide (\textit{``Authentic''}) or deepfake audio (\textit{``Deepfake''}). 
    }
    \label{tab:datasets}
\end{table*}

To assess the generalization capability of our proposed pipeline across different data distributions, we use four datasets of music recordings that contain singing voice - see Table \ref{tab:datasets}.
These include two datasets with only authentic recordings (\textsc{private} and \artisttwenty{}) and two with both authentic and deepfake tracks (\ctrsvdd{} and \wildsvdd{}). 
All recordings are resampled to \SI{16}{\kilo\hertz} for consistency.
\textbf{\textsc{private}}: A proprietary corpus of 134,826 tracks from 2,000 different singers, containing authentic commercial recordings for training \text{$\mathcal{S}$} and evaluation.
\textbf{\artisttwenty{}}~\cite{ellis2007classifying}: Open dataset of authentic recordings, used for testing only.
\textbf{\ctrsvdd{}}~\cite{zang24_interspeech}: Authentic and deepfake a cappella vocals from SVDD 2024 Challenge, used for training \text{$\mathcal{D}$} and evaluation.
\textbf{\wildsvdd{}}~\cite{zhang2024svdd}: 
Designed for real-world deepfake detection, this dataset introduced by Zhang et al. in \cite{zhang2024svdd} includes authentic and synthetic tracks sourced from social media.
Many tracks listed in the original dataset are no longer available. We focus on six artists, resulting in \num{247} tracks evenly split between authentic and fake recordings.
We use this dataset only for the evaluation of the $\mathcal{D}$ and $\mathcal{S}$ models.

For all datasets, we create an alternative version where vocals and background music are separated using \textsc{bs-roformer}~\cite{Lu2023MusicSS}, followed by non-vocal segment removal ($\pm$ 40\% per track) via an energy-based Voice Activity Detector (\textsc{vad})~\cite{sohn1999statistical}. This allows us to focus the training on samples that always contain vocals, and to use the separated background music as an augmentation. 

The model $\mathcal{D}$ uses mel-spectrogram as input, with 512 \textsc{fft} bins, 80 mel bins, a hamming window length of 400 samples, and a hop length of 160 samples.
It was trained using Binary Cross Entropy loss, a cosine annealing learning rate ($10^{-4}$ to $10^{-7}$), and $10^{-4}$ weight decay. 
Class imbalance was addressed via random oversampling, ensuring equal numbers of authentic and deepfake samples in each batch.
Singer identification ($\mathcal{S}$) models \textsc{ecapa-tdnn} and \textsc{ssl} were trained on 10-second log-mel spectrogram windows (512 \textsc{fft}, 400 window, 160 hop, 80 mel bins),  batch size of 64, early stopping (patience 10), cosine annealing learning rate ($10^{-4}$ to $10^{-7}$), and $10^{-5}$ weight decay. Data augmentation (35\% probability) included random background music, noise (impulsive/stationary)~\cite{tak2022rawboost}, and pitch shifting ($\pm2$ semitones). 

To simulate real-world conditions at inference time, we use the datasets in their original condition (i.e. no source separation or \textsc{vad}). Five \SI{10}{\second} windows are extracted from each song for model inference. For $\mathcal{D}$ we average window predictions for a final song classification. For vocalist identification models, we average the embeddings from the last dense layer of model $\mathcal{S}$, representing vocalist identity. Singer identity estimation employs cosine distance against reference embeddings, and performance is evaluated using standard speaker identification metrics (e.g., \textsc{roc}, \textsc{roc}, \textsc{eer}).

\section{Experiments \& Results}
\label{sec:results}

\begin{figure*}
    \centering
    \begin{subfigure}[b]{0.2\textwidth}
        \centering
        \includegraphics[width=\textwidth]{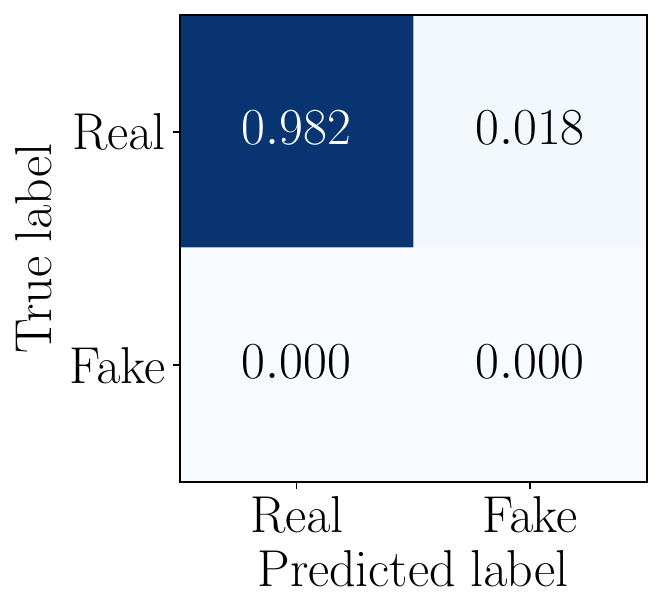}
        \caption{\textsc{private}}
    \end{subfigure}
    \hfill
    \begin{subfigure}[b]{0.2\textwidth}
        \centering
        \includegraphics[width=\textwidth]{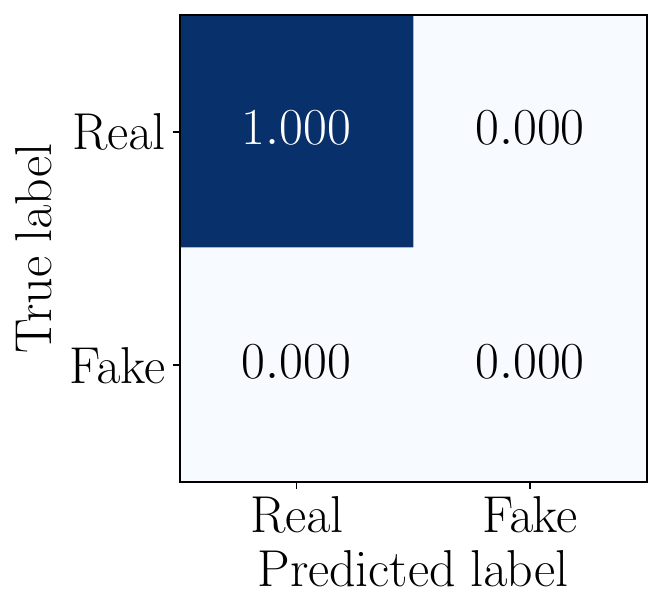}
        \caption{\artisttwenty{}}
    \end{subfigure}
    \hfill
    \begin{subfigure}[b]{0.2\textwidth}
        \centering
        \includegraphics[width=\textwidth]{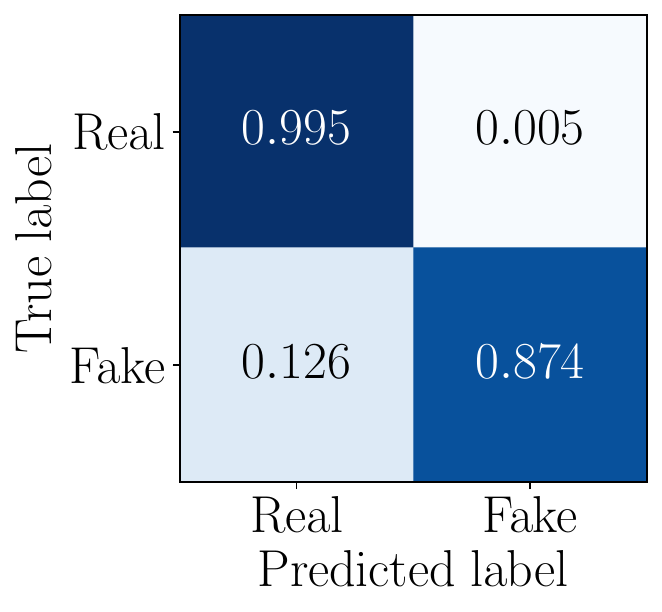}
        \caption{\ctrsvdd{}}
    \end{subfigure}
    \hfill
    \begin{subfigure}[b]{0.2\textwidth}
        \centering
        \includegraphics[width=\textwidth]{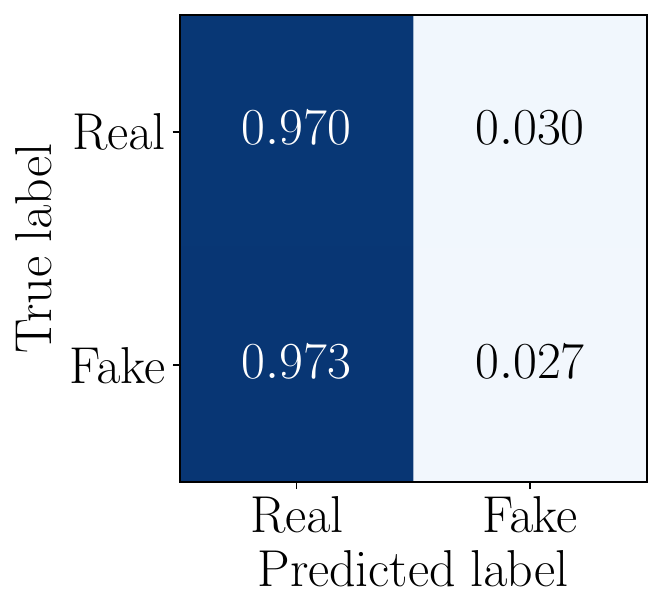}
        \caption{\wildsvdd{}}
    \end{subfigure}
    \caption{Confusion matrices of the considered singing voice discriminator $\mathcal{D}$ across the four different datasets.
    False Positive Rate (FPR) (i.e., authentic tracks misclassified as deepfakes) is exceptionally low across all datasets (see top right corners of the confusion matrices). This means that when $\mathcal{D}$  identifies a track as authentic, it is highly reliable.
    }
    \label{fig:CM_detection}
\end{figure*}

\textbf{Singer Identification:}
We first evaluate the performance of the singer identification models across all datasets.
Table~\ref{tab:datasetresults} reveals that \textsc{ecapa-tdnn} consistently outperforms baselines. 
\textsc{resnet-tdnn}'s performance is comparable to \textsc{ecapa-tdnn}'s only on \ctrsvdd{} (\textsc{auc} differences <1\%), which we attribute to the accappella content of \ctrsvdd{} being the most similar to speech. Our music-specific training regimen proves advantageous on all other datasets that contain background music. 

All models tend to exhibit poorer performance on datasets containing deepfakes.  
To investigate this further, we analyzed the \textsc{ecapa-tdnn}'s performance on the \ctrsvdd{} dataset, breaking down its effectiveness against each deepfake generation algorithm included in the corpus - see Table~\ref{tab:SV_algo}. Compelling performance is achieved on authentic data and algorithms A01-A05, A12 (\textsc{auc} > 90\%), while it degrades significantly for algorithms A07-A10 and A13 (\textsc{eer} > 30\%). 
Upon manual inspection, we observed that performance seem to correlate with the quality of the deepfakes.
The cloned voices generated by algorithms A07-A10 and A13 often do not closely resemble the original singer. This lack of fidelity likely undermines the effectiveness of the singer identification process.
It also highlights a challenge in the evaluation deepfake singer identification methods: how to deal with cases where the vocalist is not perceptually recognizable?

\begin{table*}[t]
  \centering
  \resizebox{\textwidth}{!}{ 
  \begin{tabular}{lccccccccccc}
     ~ & ~ & \multicolumn{2}{c}{Private dataset} & \multicolumn{2}{c}{Artist20} &
             \multicolumn{2}{c}{\ctrsvdd{}} & \multicolumn{2}{c}{\wildsvdd{}} &
             \multicolumn{2}{c}{Average} \\ 
             \cmidrule(lr){3-4} \cmidrule(lr){5-6} \cmidrule(lr){7-8} \cmidrule(lr){9-10} \cmidrule(lr){11-12}
             
    Model & Ref. & \textsc{eer} (\%) $\downarrow$ & \textsc{auc} (\%) $\uparrow$ & \textsc{eer} (\%)  $\downarrow$ & \textsc{auc} (\%) $\uparrow$ & \textsc{eer} (\%) $\downarrow$ & \textsc{auc} (\%) $\uparrow$ &
    
    \textsc{eer} (\%) $\downarrow$ & \textsc{auc} (\%) $\uparrow$ & \textsc{eer} (\%) $\downarrow$ & \textsc{auc} (\%) $\uparrow$ \\ \midrule
    
    \textsc{ecapa-tdnn} & \cite{desplanques2020ecapa} & \textbf{4.31}  & \textbf{98.29} & \textbf{15.56} & \textbf{91.47} & \textbf{30.34} & \textbf{76.11} & \textbf{19.24} & \textbf{87.41} & \textbf{17.36} & \textbf{88.32}        \\
  
    \textsc{ssl} & \cite{torres2023singer}  & 16.13          & 91.14          & 25.30 & 81.78          & 36.34          & 68.16          & 32.92          & 73.53 & 27.67          & 78.65          \\
  
  \textsc{resnet-tdnn} & \cite{VILLALBA2020101026}    & 8.70           & 96.28          &
  23.05          & 84.29          & 31.46          & 75.25          & 21.38
  & 86.41          & 21.15          & 85.56  \\
  \end{tabular}
  }
  \caption{Singer identification performance of three models across four different datasets.}
  \label{tab:datasetresults}
\end{table*}

\begin{table}
    \centering
    \resizebox{\columnwidth}{!}{
    \begin{tabular}{l c c c c c c c c c c c c c c}
    & A02 & \textit{REAL} & A04 & A05 & A01 & A12 & A03 & A06 & A11 & A13 & A09 & A10 & A07 & A08 \\
    \midrule
    \textsc{eer} (\%) $\downarrow$ & 8.83 & 10.73 & 11.68 & 12.01 & 13.88 & 14.16 & 14.91 & 22.99 & 24.29 & 30.36 & 33.61 & 33.98 & 36.02 & 36.05 \\
    \textsc{auc} (\%) $\uparrow$ & 96.94 & 95.48 & 95.57 & 94.21 & 93.51 & 93.53 & 91.90 & 84.63 & 83.52 & 76.33 & 71.58 & 71.12 & 68.67 & 69.17 \\
    \end{tabular}
    }
    \caption{Singer identification performance (\textsc{ecapa-tdnn}) on \ctrsvdd{} dataset. High \textsc{eer} for algorithms (A07-A10, A13) is linked to poor cloned voice fidelity, impacting identification accuracy.}
    \label{tab:SV_algo}
\end{table}

\begin{table}[H]
    \centering
        \resizebox{0.4\columnwidth}{!}{
        \begin{tabular}{@{} l l c c @{}} 
        & Pipeline & \textsc{eer} (\%) $\downarrow$ & \textsc{auc} (\%) $\uparrow$ \\
        \midrule
        \multirow{2}{*}{\ctrsvdd{}} & \textit{$\mathcal{S}$} & 30.34 & 76.11 \\
         & \textit{$\mathcal{D} \circ \mathcal{S}$} & \textbf{16.82} & \textbf{88.90} \\
        \addlinespace 
        \multirow{2}{*}{\wildsvdd{}} & \textit{$\mathcal{S}$} & 19.24 & 87.41 \\
         & \textit{$\mathcal{D} \circ \mathcal{S}$} & \textbf{15.55} & \textbf{91.55} \\
        \midrule
        \multirow{2}{*}{Average} & \textit{$\mathcal{S}$} & 24.79 & 81.76 \\
         & \textit{$\mathcal{D} \circ \mathcal{S}$} & \textbf{16.19} & \textbf{90.23} \\
        \end{tabular}
        }
    \caption{Singer identification ($\mathcal{S}$) performance of \textsc{ecapa-tdnn} with ($\mathcal{D} \circ \mathcal{S}$) and without ($\mathcal{S}$) discriminator $\mathcal{D}$. Using the discriminator significantly improves singer identification.}
    \label{tab:SV_filtered}
\end{table}

\textbf{Singing Voice Deepfake Discriminator:}
We introduced a discriminator $\mathcal{D}$ to handle poor-quality deepfakes, and evaluate it as a binary vocal deepfake classifier. The confusion matrices in Figure~\ref{fig:CM_detection} show a low \gls{fpr} across all datasets, indicating high reliability on authentic tracks.
This is an important property since, in practice, flagging an authentic track as deepfake may have damaging consequences. 
The \gls{fnr} (i.e. deepfake tracks misclassified as authentic) is significantly higher in \wildsvdd{} than in \ctrsvdd{}. Given our empirical observation of the varying deepfakes quality in \ctrsvdd{}, we hypothesize it  may explain the difference in \gls{fnr}. \wildsvdd{}'s higher \gls{fnr} would then suggest it contains higher-quality deepfakes that can effectively fool $\mathcal{D}$, whereas \ctrsvdd{}'s lower quality deepfakes are easier to detect.

\textbf{Combining Singing Voice Deepfake Detection and Singer Identification:}
As a final experiment, we evaluate the benefits of the proposed 2-step pipeline described in Figure \ref{fig:pipeline}.
We evaluate the framework on the \ctrsvdd{} and \wildsvdd{} datasets, considering only tracks classified as authentic by the discriminator (we label this condition $\mathcal{D} \circ \mathcal{S}$) and compare it to the performance of the singer identification model only (labeled $\mathcal{S}$).
Table \ref{tab:SV_filtered} shows the results for \textsc{eer} and \textsc{auc} metrics, which reveal that introducing $\mathcal{D}$ significantly enhances singer identification performance. Combining the results of our experiments and our empirical observation of varied deepfake quality, our interpretation is that ensuring $\mathcal{S}$ operates on realistic vocal likenesses, makes the singer identification task more meaningful and tractable (attempting to identify a singer in the case of an unidentifiable likeness is bound to fail).
In future work we propound to cross-reference these results with a study of deepfakes perceptual quality to test our interpretation further. 

\section{Conclusions}
\label{sec:conclusion}

This paper addresses singer identification in authentic and deepfake singing voices recordings. We proposed a novel two-stage pipeline based on the premise that the highest quality deepfakes are those with the greatest potential for harm. 
Our experiments show that our proposed method outperforms baselines on multiple benchmarks. Combining these results with empirical observation suggests that the performance of singer identification models degrades on low quality deepfakes, where the vocal likeness is not faithfully reproduced. Our interpretation is that the introduction of the discriminator allows the singer identification model to only operate on high quality deepfakes and therefore makes the identification task more meaningful and tractable. 
For future work we recommend a perceptual study of deepfake quality to further test this interpretation.

\bibliographystyle{unsrt}
\bibliography{bibliography.bib}

\end{document}